\documentclass{article}
\usepackage{ijcai19}

\usepackage{times}
\usepackage{soul}
\usepackage{url}
\usepackage[hidelinks]{hyperref}
\usepackage[utf8]{inputenc}
\usepackage[small]{caption}
\usepackage{graphicx}
\usepackage{amsmath}
\usepackage{booktabs}
\usepackage{algorithm}
\usepackage{algorithmic}
\usepackage{xcolor}
\usepackage{subfigure}
\title{ES-CTC: A Deep Neuroevolution Model for Cooperative Intelligent Freeway Traffic Control }

\author{
Yuankai Wu$^1$\and
Huachun Tan$^{2*}$\and
Zhuxi Jiang$^{3}$\And
Bin Ran$^4$
\affiliations
$^1$School of Mechanical Engineering, Beijing Institute of Tachnology, China\\
$^{2*}$School of Transportation Engineering, Southeast University, China\\
$^3$Momenta, China\\
$^4$College of Engineering, University of Wisconsin-Madison, USA
\emails
\ $^1$Kaimaogege@gmail.com,
$^{2*}$tanhc@seu.edu.cn,
$^{3}$zjiang9310@gmail.com,
$^{4}$bran@engr.wisc.edu,
}

\begin{document}

\maketitle

\begin{abstract}
Cooperative intelligent freeway traffic control is an important application in intelligent transportation systems, which is expected to improve the mobility of freeway networks. In this paper, we propose a deep neuroevolution model, called ES-CTC, to achieve a cooperative control scheme of ramp metering, differential variable speed limits and lane change control agents for improving freeway traffic. In this model, the graph convolutional networks are used to learn more meaningful spatial pattern from traffic sensors, a knowledge sharing layer is designed for communication between different agents. The proposed neural networks structure allows different agents share knowledge with each other and execute action asynchronously. In order to address the delayed reward and action asynchronism issues, the evolutionary strategy is utilized to train the agents under stochastic traffic demands. The experimental results on a simulated freeway section indicate that ES-CTC is a viable approach and outperforms several existing methods.
\end{abstract}

\section{Introduction}
The ongoing drastic expansion of car ownership and travel demand have led to increasing freeway congestion, with adverse effects on the economy. To relieve freeway congestion, numerous freeway traffic control approaches, e.g. dynamic routing, variable speed limit (VSL), ramp metering (RM), lane change control (LCC) etc., are studied. From a systematic viewpoint, using one management approach alone cannot fully optimize the freeway traffic in practice. The mainlane flow, on-ramp flow, routing behaviors and lane changing behaviors need to be regulated in a coordinated manner in order to improve the freeway condition. This is the motivation for investigating the coordination of different traffic control approaches. 

There is a large volume of published studies describing the cooperative traffic control: Hedgy et.al~\shortcite{hegyi2005model} developed a predictive coordinated control approach for the coordination of VSL and RM. Carlson et.al~\shortcite{carlson2010optimal} formulated coordinated VSL and RM control as an optimal control problem using second-order traffic flow model. Recently, the coordination of RM, VSL and LCC under connected autonomous vehicle environment was studied~\cite{roncoli2015traffic}. Two limitations worth noting in respect of the studies mentioned above are: 1) The control model are highly dependent on the integrated traffic flow models, which are inevitably inconsistent with the real-world traffic breakdown. 2) The success of proactive approaches are based on robustness and reliability of the short-term traffic
prediction model. The accurate and reliable short-term traffic prediction is not an easy task because the evolution of traffic state is related to many factors~\cite{wu2018hybrid}.  

Recently, the advent of deep reinforcement learning (DRL) has lead to potential applications of reinforcement learning (RL) techniques to tackle challenging control problems in intelligent transportation systems. DRL has given promising results in RM~\cite{belletti2018expert}, traffic light control~\cite{wei2018intellilight}, differential VSL control~\cite{wu2018differential}, fleet management~\cite{lin2018efficient} and hybrid electric vehicle energy management~\cite{wu2018continuous}. The utilization of deep learning algorithms within RL allows a well-trained traffic control agent achieves a proactive control scheme, and optimizes the transportation benefits. The success of DRL on one specific traffic control approach hold great promise for application of DRL on coordination of different traffic control approaches.

However, the coordination of different traffic control approaches within one DRL framework is not an easy task. The first challenge is due to the difference between the control cycle of different agents. In many situations, the agents change actions asynchronously, a somewhat different situation from that familiar from popular multi-agent DRL frameworks~\cite{foerster2016learning,lowe2017multi}. For example, the agents controlling on-ramp flow should decide whether to change traffic light phase every few seconds. While the control cycle for VSL agents are always above 1 minute because a frequently change speed limit will unstablilize the traffic flow.

The second challenge stems from the difficulties in defining a representative reward signal for different traffic control agents. The aim of traffic management would be to reduce travel time and increase traffic flow. However, the average travel time and total flow cannot be computed until all the vehicles have completed their routes, which causes the issue of delayed rewards~\cite{van2016coordinated}. The delayed rewards would cause further credit assignment problems in multi-agent DRL~\cite{foerster2017counterfactual}. 

The third challenge lies in the modeling of the traffic state. Traditional, the traffic state collected from sensors are modeled as images and/or vectors, and is directly taken as an input for a convolutional neural networks (CNN)~\cite{wei2018intellilight} or fully connected neural networks (FC)~\cite{li2016traffic}. However, sensors on the road network contain complex spatial correlations and exhibits graph structure. There have been numerous studies reported that the graph convolutional network (GCN) is more suitable for modeling spatial correlation of traffic sensors than CNN and FCN in traffic prediction~\cite{li2018diffusion,lv2018lc}.  

To tackle those challenges, we propose a deep neuroevolution~\cite{salimans2017evolution} based multi-agent framework for cooperative traffic control (ES-CTC). The main contributions of this paper can be summarized as follows:
\begin{enumerate}
\item We find that the deep neuroevolution approach is a perfect match for cooperative traffic control. In deep neuroevolution approach like evolutional strategies (ES), the only feedback signal for different agents is the final return of an episode. As a result, the problem of delayed reward is readily solved with ES. 
\item We proposed a novel structure named knowledge sharing graph convolutional nets (KS-GCN) to generate control actions from state collected from traffic sensors. GCN is used as the building block for the proposed structure, which can fully capture the spatial dependency between different sensors. The structure allows communication and knowledge-sharing between different agents. Based on the knowledge sharing layer, the neural agent can coordinate with other agents by executing action in its own control circle.
\item The travel demands for training the neural networks are modeled as a stochastic distribution, leading to the changes in system dynamics of the environment. The experiments show that the proposed approach works well under stochastic travel demands.
\end{enumerate}

\section{Problem Statement} 
\label{PS}
\begin{figure}[h]
\begin{center}
\includegraphics[width=0.5\textwidth]{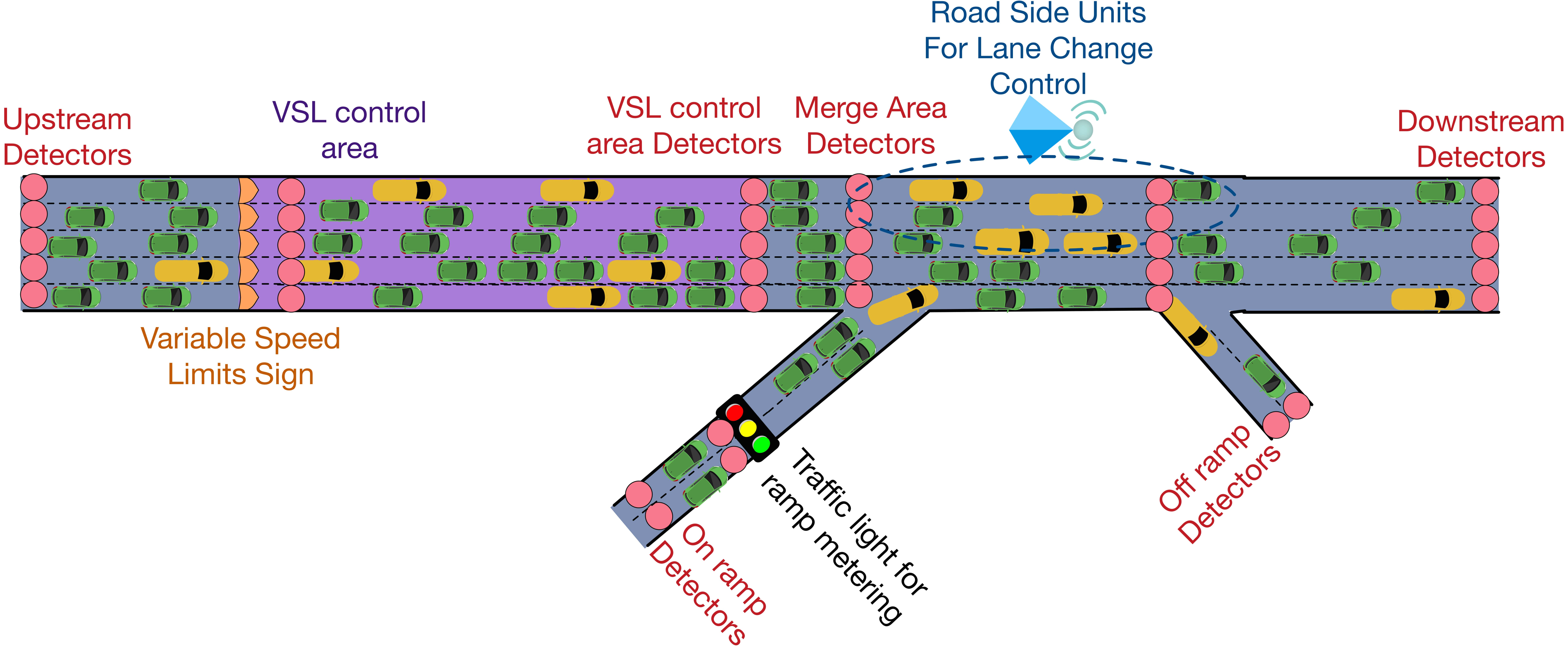}
\caption{The freeway section has an on-ramp and an off-ramp. There is a recurrent bottleneck caused by conflicts between inflow in on-ramp and outflow in mainlane. There are several traffic detectors and controllers in this freeway section. The controllers include VSL signs, traffic lights for RM and road side units for LCC. }
\label{problemstate}
\end{center}
\end{figure}

The freeway section considered in this paper is given in Figure \ref{problemstate}. The freeway section in Figure \ref{problemstate} is composed by multiple lanes and it presents an on-ramp and an off-ramp. As it may be seen in the figure, the interference between vehicles is appearing in the merging area between inflow of on-ramp and outflow of mainstream. The conflicts cause further speed reductions in the merging area, contributing to the creation of a generalised bottleneck. 

Following the statement in~\cite{roncoli2015traffic}, we consider that the freeway flow with a high ratio of connected autonomous vehicle (CAV). Therefore the differential VSL and LCC can be successfully implemented. More specifically, the following control agents are considered in this paper:
\begin{enumerate}
\item[•] \textbf{Ramp-metering agent:} The agent is to regulate the inflow from on-ramp to mainstream by change the phase of the traffic light in on-ramp.
\item[•] \textbf{Differential VSL (DVSL) agent:} The DVSL agent aims at regulating the outflow of controlled area to prevent the capacity drop at bottlenecks. The conflicts between vehicles occur mostly in the right lanes. Therefore different speed limits among lanes might be more effective. The DVSL strategy can be implemented under CAV environment. The DVSL signs can send speed limit orders to the vehicles in the corresponding lane, the vehicles are forced to drive under the received speed limit.
\item[•] \textbf{LCC agent:} The LCC is used to regulate the lateral flows for each lane. The implementation of LCC agent is more challenging than RM and DVSL agents. In this paper, we only considered to use a road-side unit (RSU) to send ``keep lane'' orders to the vehicle in left 2 lanes of the merge area. The reason is that the lateral inflow from left lanes to right lanes will cause severe congestion when traffic breakdown occured in the merge area of the right lanes. 
\end{enumerate}
Each control agent executes its own action according to its own control cycle. We denote by $T^{R}$ the control cycle for RM agent, $T^{D}$ for DVSL agent and $T^{L}$ for LCC agent. The main goal of these agents is to reduce congestion and promote the freeway capacity in a coordinated manner. 

\section{The KS-GCN Model Description}

\begin{figure}[h]
\begin{center}
\includegraphics[width=0.42\textwidth]{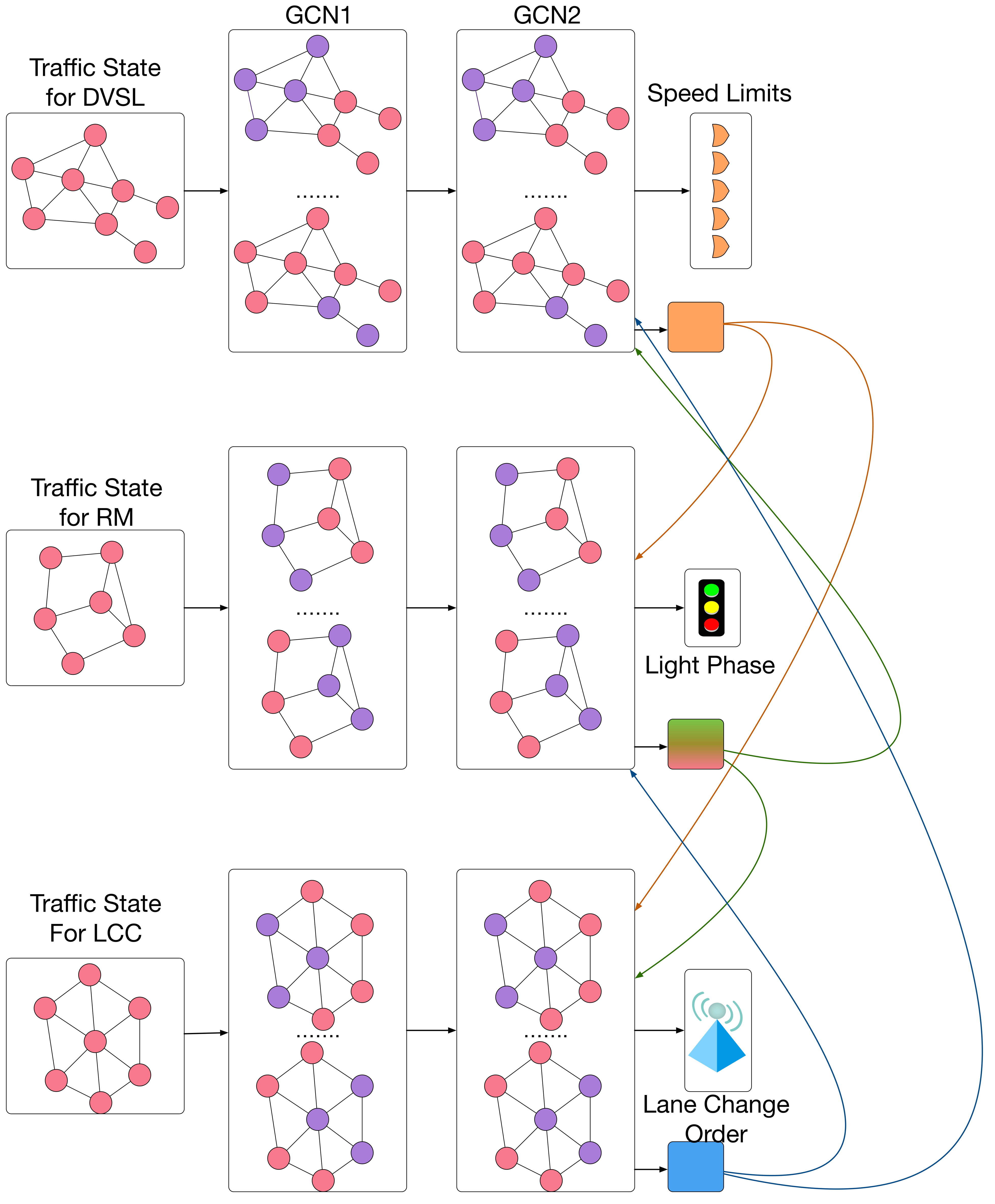}
\caption{The architecture of KS-GCN }
\label{ksgcn}
\end{center}
\end{figure}

Figure \ref{ksgcn} presents the architecture of KS-GCN, which is comprised of several GCN layers, traffic state inputs for DVSL, RM, LCC, several knowledge sharing layers, DVSL, RM and LCC actuators respectively.

\subsection{Framework}
The function of KS-GCN is to generate coordinated actions for the DVSL agent, RM agent and LCC agent given observed traffic state from correlated sensors/detectors on the targeted freeway section. Each agent only receives states from its mostly related sensors. Each sensor collects $P$ traffic variables (e.g., velocity, occupancy rate) in one cycle and is denoted as a vector $x_{T-T^c} \in R^P$. The sensor network can be represented as a weighted undirected graph $\mathcal{G} = (\mathcal{V},\mathcal{E},\mathbf{W})$, where $\mathcal{V}$ is a set of nodes $|\mathcal{V}| = N$, $\mathcal{E}$ is a set of edges, $\mathbf{W} \in R^{N \times N}$ is a weighted adjacency matrix. 
The KS-GCN learns functions that map graph signals to traffic control signals asynchronously:
\begin{equation}
\footnotesize
\begin{split}
[\mathbf{X}^{R}_{T-T^{R}}, \mathbf{X}^{D}_{T-T^{D}}, \mathbf{X}^{L}_{T-T^{L}}, \mathbf{W}^R, \mathbf{W}^D, \mathbf{W}^L] \to \mathbf{a^{RM}},\\
 if \quad T = iT^{R}\\
[\mathbf{X}^{R}_{T-T^{R}}, \mathbf{X}^{D}_{T-T^{D}}, \mathbf{X}^{L}_{T-T^{L}}, \mathbf{W}^R, \mathbf{W}^D, \mathbf{W}^L] \to \mathbf{a^{DVSL}}, \\
 if \quad T = iT^{D}\\
[\mathbf{X}^{R}_{T-T^{R}}, \mathbf{X}^{D}_{T-T^{D}}, \mathbf{X}^{L}_{T-T^{L}}, \mathbf{W}^R, \mathbf{W}^D, \mathbf{W}^L] \to \mathbf{a^{LCC}}, \\
 if \quad T = iT^{L}
\end{split}
\end{equation}
where $\mathbf{X}^{R} \in R^{N^R \times P}$, $\mathbf{X}^{D} \in R^{N^D \times P}$ and $\mathbf{X}^{L} \in R^{N^L \times P}$ are graph sensor signals that related to RM, DVSL and LCC agents respectively.The 3 agents can share sensors, therefore $N^R + N^D + N^L \geq N$. $\mathbf{W}^R$, $\mathbf{W}^D$ and $\mathbf{W}^L$ are RM, DVSL and LCC similarity matrices derived from $\mathbf{W}$. $i$ is an integer. KS-GCN asynchronously updates the control signals every control cycle. The control cycles of RM ($T^{R}$), DVSL ($T^{D}$) and LCC ($T^{L}$) can be different from each other.

\subsection{Network Structure}
We use the GCN architecture proposed in \cite{kipf2016semi} to learn the spatial dependence between traffic signals on the graph. The layer-wise propagation rule of the specific GCN is:
\begin{equation}
\mathbf{H}^{(l+1)} = \alpha^l(\bar{\mathbf{D}}^{-\frac{1}{2}} \bar{\mathbf{W}} \bar{\mathbf{D}}^{-\frac{1}{2}}\mathbf{H}^{(l)}\mathbf{U}^{(l)} + \begin{pmatrix} b^{(l)} \\
. \\
. \\
. \\
b^{(l)} \end{pmatrix})
\end{equation} 
where $\bar{\mathbf{W}} = \mathbf{W} + I_N$ is the adjacency matrix that added self-connections. $I_N$ is the identity matrix. $\bar{\mathbf{D}}_{ii} = \sum_j \bar{\mathbf{W}}_{ij}$. $\mathbf{U}^{(l)} \in R^{f^{(l)} \times f^{(l+1)}}$, $b^{(l)} \in R^{f^{(l+1)}}$ are the layer-specific trainable weight matrix and bias. $\mathbf{H}^{(l)} \in R^{N \times f^{(l)}}$, $N$ is the number of graph signal, $f^{(l)}$ is the number of feature in $l$-th layer, and $\alpha^l()$ is the activation in $l$-th layer. In KS-GCN, there are 3 stacked GCNs, which are used to learn features from traffic states for RM, DVSL and LCC agents respectively.

On top of the GCN, we further use a knowledge sharing layer to learn the sharing features for each agent. After $L$ layers of GCN, the last output matrix $\mathbf{H}^{L}$ is of size ${N \times f^{(L)}}$. We use a simple FC layer for knowledge sharing, the output matrix is reshaped as a vector $h^{L} \in R^{Nf^{(L)}}$. The sharing feature $s$ can be obtained by: 
\begin{equation}
s = \alpha^{ks}(\mathbf{U}^{ks}h^{L}+b^{ks}),
\end{equation}  
$\mathbf{U}^{ks} \in R^{K \times Nf^{(L)}}$ and $b^{ks} \in R^K$ are trainable weights for the knowledge sharing layer. $K$ is the dimension of the sharing knowledge. Each agent shares its own knowledge with the other agents for generating specific action. The sharing process is done by concatenation:
\begin{equation}
\begin{split}
z^{RM} = concat(h^{L,RM}, s^{DVSL}, s^{LCC}), \\
z^{DVSL} = concat(h^{L,DVSL}, s^{RM}, s^{LCC}), \\
z^{LCC} = concat(h^{L,LCC}, s^{RM}, s^{DVSL}).
\end{split}
\end{equation}
Here, $z$ is the final vectorized feature for generating control action, $concat$ is the concatenation layer. 

\subsection{Action Design}
\label{AD}
In this subsection, we introduce the action representation of different agents. The action for RM is represented by the phase of traffic light in the on-ramp. It is defined as $\mathbf{a}^{RM} = 1$: change the light to green phase (the vehicles in on-ramp is allowed to enter the freeway), and $\mathbf{a}^{RM} = 0$: change the light to red phase. The action for RM agent can be generated by a FC layer with softmax activation:
\begin{equation}
\mathbf{a}^{RM} = argmax(softmax(\mathbf{U}^{RM}z^{RM} + b^{RM}))
\end{equation}
where $\mathbf{U}^{RM} \in R^{2 \times f^{RM}}$, and $b^{RM} \in R^{2}$ are the trainable weights. $argmax$ is used to find the index with maximum value. 

A similar action design can be applied to LCC agent. The action of LCC agent is defined as $\mathbf{a}^{LCC} = 1$: allow lane change in left 2 lanes, and $\mathbf{a}^{LCC} = 0$: forbidden lane change in left 2 lanes. The generation process of $\mathbf{a}^{LCC}$ is:
\begin{equation}
\mathbf{a}^{LCC} = argmax(softmax(\mathbf{U}^{LCC}z^{LCC} + b^{LCC})).
\end{equation}

The action $\mathbf{a}^{DVSL}$ interacts the speed limit of all lanes in the controlled area. Therefore $\mathbf{a}^{DVSL} \in R^{c}$, where $c$ is the number of lane at the controlled section. Considering the real world implementation and the driver compliance issue, the elements of $\mathbf{a}^{DVSL}$ is set as discrete values $\mathbf{a}^{DVSL}_i \in [0,1,\cdots,M]$. And the speed limits $V \in R^{c}$ is equal to $V_0 + j\mathbf{a}^{DVSL}$, where $V_0$ is the minimum value of the speed limit, $j$ is the integer multiples, the maximum value of speed limits is $V_0 + jM$. It is not feasible for a neural networks to generate explicit discrete speed limits for multiple lanes because the total number of actions for a $c$-lane freeway section will be as large as $M^c$. The neural networks with limited size is difficult or impossible to handle such a large action
space. Follow the work in \cite{wu2018differential}, the action generation process for the DVSL agent is defined as:
\begin{equation}
\mathbf{a}^{DVSL} = int((M+1)sigmoid(\mathbf{U}^{DVSL}z^{DVSL} + b^{DVSL})),
\label{Mspeed}
\end{equation}
The activation of FC layer for DVSL agent is $sigmoid$ function. The outputs of the FC layer are then multiplied with $M+1$. The discrete action $\mathbf{a}^{DVSL}$ is obtained by the integer parts of the scaled outputs.

\section{Evolutionary Strategy for Optimization}
In this section our aim is to propose an efficient and effective optimization algorithm for coopetative traffic control using KS-GCN based on evolutionary strategy (ES). Finding an optimal coopetative control policy for a given freeway section in section \ref{PS} can be seen as an optimization problem to search for a trainable parameter set $\theta$ for KS-GCN that maximize the total outflow $F(\theta) = \sum^T_0 r_t$ of the freeway section. $r_t$ is the instantaneous outflow of the freeway section.

The parameters $\theta$ of KS-GCN can be directed updated by using the final return $F_j(\theta + \sigma\epsilon_j)$ of parallel workers in ES, therefore we proposed to use ES as the optimization algorithm for KS-GCN. Another objective of the freeway control agents is to achieve an optimal control scheme under stochastic traffic demand. This also can be easily done via ES. In simulation, the traffic demand is modeled as a random process. In each episode, a new traffic demand is set by sampling data from the random process, then several parallel workers are used to run on simulations with the same traffic demand, finally the parameters $\theta$ is updated by the final returns of these parallel workers. We find that this stochastic training approach guarantees the generalization of the agents. 

Another core challenge is how to balance exploration and exploitation using ES. The total outflow as the reward function is sometimes deceptive, e.g, the agents that achieved high outflow for a specific traffic demand might perform badly under another traffic demand sampled from the same random process. Without adequate exploration, the agents might fail to discover effective traffic control strategies. In this paper, we exploit the novelty-seeking (NS) proposed in \cite{conti2018improving} for exploration. In NS, the novelty of one policy is characterized by a behavior vector $b(\pi_\theta)$ that describes its behavior. For CTC, we define traffic demand $D$ specific $b(\pi_\theta, D)$ as:
\begin{equation}
b(\pi_\theta, D) = |avg(\mathbf{a}_{D}^{RM}), avg(\frac{\mathbf{a}_{D}^{DVSL}}{M+1}), avg(\mathbf{a}_{D}^{LCC})|
\end{equation} 
where $\mathbf{a}_{D}^{RM}$, $\mathbf{a}_{D}^{DVSL}$ and $\mathbf{a}_{D}^{LCC}$ are vectors that contain all time RM, DVSL and LCC actions under demand $D$. The original work of NS use a set of parameters to calculate the novelty. Because the traffic demand changes every episode, calculating demand specific behavior vectors for a set of parameters will be very time-consuming. In this paper, the novelty of a parallel worker is directly defined as the distance between its behavior vector and the one of unperturbed agent on demand $D$:
\begin{equation}
N_j(b(\pi_{\theta+\sigma\epsilon_j}, D), b(\pi_{\theta}, D)) = \|b(\pi_{\theta+\sigma\epsilon_j}, D)-b(\pi_{\theta}, D) \|_2,
\end{equation}
The parameter update rule for ES-CTC is then expressed as follows:
\begin{equation}
\theta_{t+1} = \theta_{t} + \alpha\frac{1}{n\sigma}\sum^n_{j=1} (1-w)F_j + wN_j
\label{novelty}
\end{equation}
where $n$ is the number of parallel workers, $\alpha$ is the learning rate. $0\leq w \leq 1$ is the parameter to balance between exploration and exploitation. In this work, we slowly decrease $w$ every episode. Algorithm \ref{es-ctc} summarizes the optimization procedure of ES-CTC

\begin{algorithm}[tb]
\caption{ES-CTC}
\label{es-ctc}
\textbf{Input}: Learning rate $\alpha$, noise standard deviation $\sigma$, random demand procees $P_D$, balance parameter $w$
\begin{algorithmic}[1] %[1] enables line numbers
\STATE \textbf{For} {t = 1,2,...} \textbf{do} \\
\STATE $\quad$  Sample traffic demand $D_t$ from $P_D$ \\
\STATE $\quad$  Compute bahavior vector $b_t(\theta_t, D_t)$ \\
\STATE $\quad$  \textbf{For} each work $j$ = 1,2,...,n \textbf{do} \\
\STATE $\quad$ $\quad$    Sample $\epsilon_j \sim N(0, I)$ \\
\STATE $\quad$ $\quad$    Compute returns $F_j$ and novelties $N_j$ using Eq (\ref{novelty})\\
\STATE $\quad$  \textbf{End For} \\
\STATE $\quad$   Set $\theta_{t+1} = \theta_{t} + \alpha\frac{1}{n\sigma}\sum^n_{j=1} (1-w)F_j + wN_j$\\
\STATE $\quad$   Decrease $w$ \\
\STATE \textbf{End For} \\
\end{algorithmic}
\end{algorithm}

\section{Experiments}
In this section, we mainly conducted experiments on a simulated freeway section built by SUMO to evaluate the effectiveness of ES-CTC.
\subsection{The simulated freeway section}
\begin{figure}[h]
\begin{center}
\includegraphics[width=0.45\textwidth]{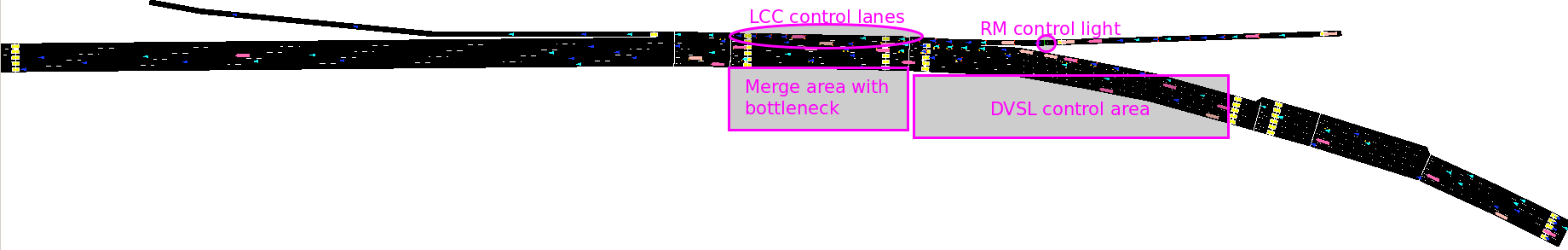}
\caption{The freeway section in SUMO }
\label{sumoen}
\end{center}
\end{figure}

The open source software SUMO is selected for the experiments. The software supports set the speed limits for each lane, set traffic phase for traffic light and forbidden lane changing using its API–the Traffic Control Interface (TraCI) package. A 874.51m freeway section with on- and off- ramps of I405 north bound in California, USA is selected. The original speed limits for the mainlane of this section are $65mile/h$, for the on- and off- ramps are $50mile/h$. The freeway section in SUMO and each agents' control area can be found in Figure \ref{sumoen}. The travel demand of this freeway can be categorized into 3 routes: 1) From mainlane to mainlane (M2M), 2) From mainlane to off-ramp (M2Off), and 3) From on-ramp to mainline (On2M). Based on observation from recorded traffic flow from sensors of PeMS\footnote{http://pems.dot.ca.gov}, the hourly demand of these 3 routes is modeled as Poisson distribution with average value 5427, 1809 and 1153 respectively. The depart lane of the vehicles are randomly set according to uniform distribution. Passenger car with a length 3.5m and truck/bus with a length 8m are selected as vehicle types in the simulated traffic stream. The type of vehicles are selected randomly according to probability $[0.85, 0.15]$. Each round simulation lasts for 1 hour. 

We place sensors in the upstream of DVSL controlled area, DVSL controlled area, on-ramp and merge area to detect the traffic state. The sensors on off-ramp and downstream area are used to calculate the outflow of the freeway section. The outflow can be used to compute the final return for the agents. The traffic speed and occupancy rate collected from these sensors are used as inputs for the KS-GCN. Specifically, the on-ramp and upstream of merge area are used for RM agent. The sensors in the upstream of DVSL controlled, DVSL controlled area and upstream of merge area are used for DVSL agent. The sensors in the merge area are used for LCC agent. The sizes of $\mathbf{X}^{R}$, $\mathbf{X}^{D}$ and $\mathbf{X}^{L}$ are $8 \times 2$, $22 \times 2$ and $12 \times 2$ respectively. The element $w_{ij}$ of similarity matrix $\mathbf{W}$ for input states is given by:
\begin{equation}
w_{ij} = \begin{cases}
exp^{\frac{-|loc(i)-loc(j)|}{10}} \quad if \quad (i,j) \in \mathcal{D} \\
0.9 \quad if \quad (i,j) \in \mathcal{S} \\
1 \quad if \quad i =j
\end{cases}
\end{equation}
where $loc()$ denotes the location of the sensor. $(i,j) \in \mathcal{D}$ means that sensor $i$ and sensor $j$ belong to different freeway sections. $(i,j) \in \mathcal{S}$ denotes that sensor $i$ and sensor $j$ are in the same freeway section. The control cycle $T^{R}$, $T^{D}$ and $T^{L}$ of RM, DVSL and LCC agents are set to 3, 60 and 30 seconds respectively. The speed limits set for DVSL agent is $[10mph, 15mph,\cdots, 75mph]$.

\subsection{Benchmarks} 
We compare ES-CTC with the following baseline methods, which include numerous DRL based traffic control models:
\begin{itemize}
\item[•]\textbf{No control:} The baseline without any DVSL, RM and LCC control.
\item[•]\textbf{DQN-RM} A modified version of DQN based traffic light control for RM. The state input of the neural networks is the vectorization of $\mathbf{X}^{R}$. The agent is modeled as a neural networks with two hidden FC layers.
\item[•]\textbf{TRPO-RM} The actor and critic of the agent are modeled as neural networks with two hidden FC layers.
\item[•]\textbf{DDPG-DVSL} A DRL based DVSL control model whose actor and critic of the agent are modeled as a neural networks with two hidden FC layers.
\end{itemize}

The traffic state $\mathbf{X}^{R}$ is used as the state variable for DQN-RM and TRPO-RM. The traffic state $\mathbf{X}^{D}$ is used as the state variable for DDPG-DVSL. The neural networks of DQN-RM, actor and critic of DDPG-DVSL and TRPO-RM have 2 hidden FC layers, which contain 30 hidden neurons and 20 hidden neuron respectively. The agents of ES-CTC are built upon 2 layer GCNs, the numbers of feature in 1st and 2nd are 5 and 3 respectively, the dimensions of sharing feature are set as 8. The reward signal of DQN-RM, TRPO-RM and DDPG-DVSL is the outflow $r_t$ of the freeway section at time point $t$. Their discount factors are set to 0.9. The return $F_j$ for ES-CTC is the total outflow of the freeway section.

\subsection{Performance Comparisons}
\subsubsection{Scenario 1}
\begin{figure}[htb]
  \centering
  \subfigure[Evolution of total outflow of DQN-RM, TRPO-RM and DDPG-DVSL]{
    \label{DRLs1} %% label for first subfigure
    \includegraphics[width=0.4\textwidth]{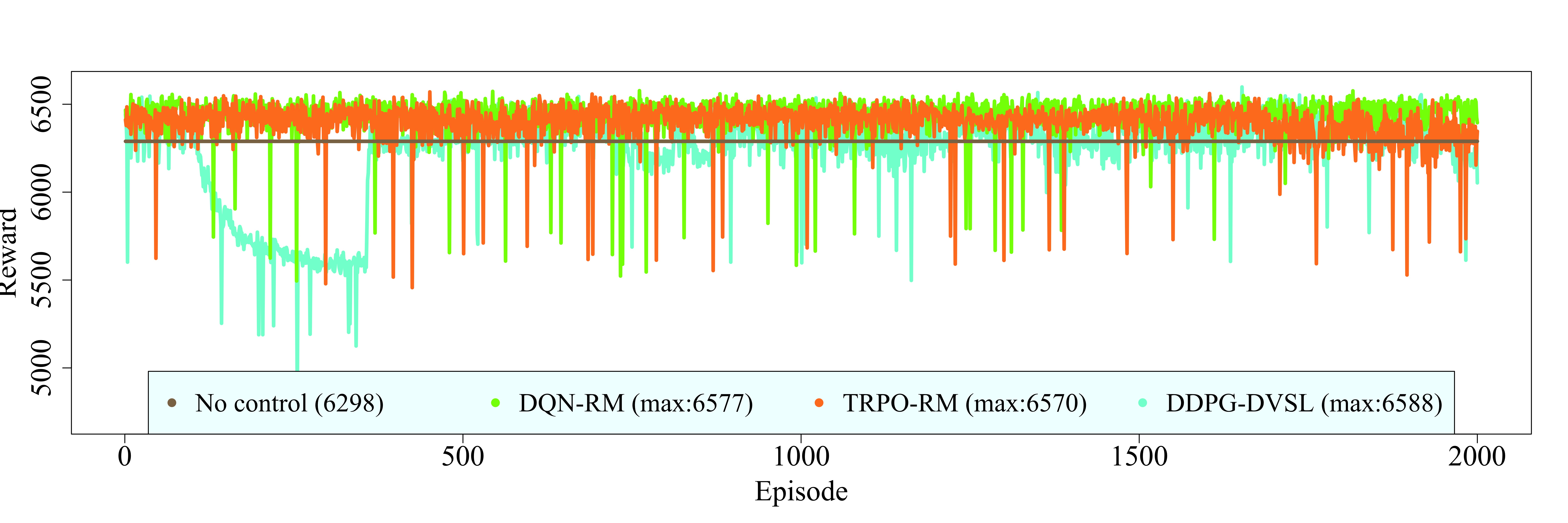}}
  \hspace{0.0in}
  \subfigure[Evolution of total outflow of the ES-CTC model]{
    \label{DNs2} %% label for first subfigure
    \includegraphics[width=0.4\textwidth]{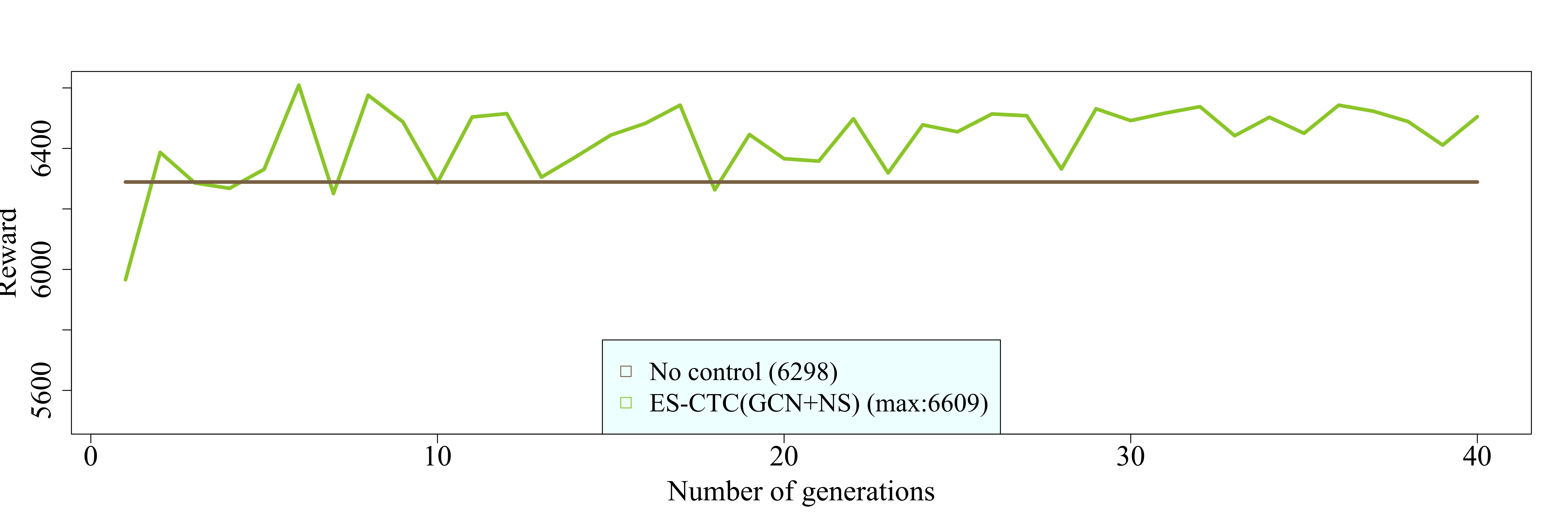}}
  \hspace{0.0in}
  \caption{Evolution of total outflow of the models over iterations of the algorithms.}
  \label{ts1} %% label for entire figure
\end{figure}

We first evaluate all models on a simple case, they are constantly optimized on a same demand profile. The DRL based DQN-RM, TRPO-RM and DDPG-DVSL are trained with the demand with 2000 episodes.  The number of parallel workers $n$ for ES-CTC is set to 50. To make the comparison fair, we update the parameters of ES-CTC 40 times therefore all models are learned with same number of simulation.  In this scenario, we can observe whether the compared models can converge to a stable and optimal point by the training process of all models. The evolution of the overall outflow of each algorithm during training can be seen in Figure \ref{ts1}.

We discover that the DQN-RM, TRPO-RM and DDPG-DVSL fail to converge to a stable value. Several oscillations can be observed from Figure \ref{DRLs1}. The outflow are related to many other factors such as the inflow of on-ramp and outflow of off-ramp, which could not fully controlled by the agents. Moreover, the vehicle can be computed as a out vehicle only when it has leaved the freeway section, there could be a delay between the control effects of the agents on the vehicle and computation of reward signal. These issues make the DRL based approaches difficult to converge. It is observed that ES-CTC is more stable from Figure \ref{DNs2}. ES-CTC reaches a relatively high outflow after 25 round generation and achieves the highest max outflow with 6609 vehicles. Another advantage of ES-CTC models is that they are significantly faster than DRL models due to their higher parallelization capability. The results indicate that deep neuroevolution model is more suitable for cooperative traffic control compared with DRL models. The total number of outflow only reaches 6289 when no control strategy is implemented. The maximum outflows of all DRL models and ES-CTC are significantly higher than 6289. The maximum outflows for DQN-RM, TRPO-RM and DDPG-DVSL are 6577, 6570 and 6588 respectively. It shows that the traffic control strategies can promote the capacity of the freeway. 
\subsubsection{Scenario 2}   
In the second case, the DQN-RM, TRPO-RM, DDPG-DVSL and ES-CTC are trained and evaluated on stochastic traffic demand. The DRL based DQN-RM, TRPO-RM and DDPG-DVSL are trained with the demand with 3000 episodes. They are trained with a new traffic demand in each episode. The number of parallel workers $n$ for ES-CTC is set to 100. In order to guarantee all models consume similar wall-clock time, we evolved the ES-CTC model with 200 generations. After training, we compare the average outflow of all models on 100 stochastic demands. The traditional performance metric used in the RL problems is the average total return achieved by the model in an episode. In order to obtain more representative metrics independent of reward shaping for traffic control, we also compute the average traffic demand satisfaction degree $TDS$ and average improvement level $IL$, which are defined as
\begin{equation}
\begin{split}
TDS = \frac{F_i}{D_i} \\
IL = \frac{F_i - F^{N}_i} {F^{N}_i}.
\end{split}
\end{equation}      
Here $D_i$ is the total demand of the $i$th episode, $F^{N}_i$ is the total outflow of $i$th episode without any traffic control agents. The evaluation results of 4 models are given in Table \ref{tab:plain}. We can find ES-CTC achieves relatively higher average outflow, $TDS$ and $IL$ than three DRL benchmarks on 100 stochastic traffic demands. The ES based optimization strategy, graph convolutional structure and coordination between different agents are the keys to its success. 

\begin{table}
\centering
\begin{tabular}{cccc}
\toprule
Models  & Outflow & $TDS$ & $IL$ \\
\midrule
ES-CTC  & \textbf{6725.1}  & \textbf{0.7949}  &\textbf{0.0656} \\
DQN-RM  & 6567.2  & 0.7819  &0.0471  \\
TRPO-RM & 6563.9  & 0.7839  &0.0437  \\
DDPG-DVSL &6642.3 & 0.7904  &0.0502  \\
\bottomrule
\end{tabular}
\caption{The average evaluation metrics on 100 stochastic traffic demands}
\label{tab:plain}
\end{table}

\begin{figure}[htb]
  \centering
  \subfigure[The RM action produced by ES-CTC in the first 6 minute]{
    \label{DRLs1} %% label for first subfigure
    \includegraphics[width=0.2\textwidth]{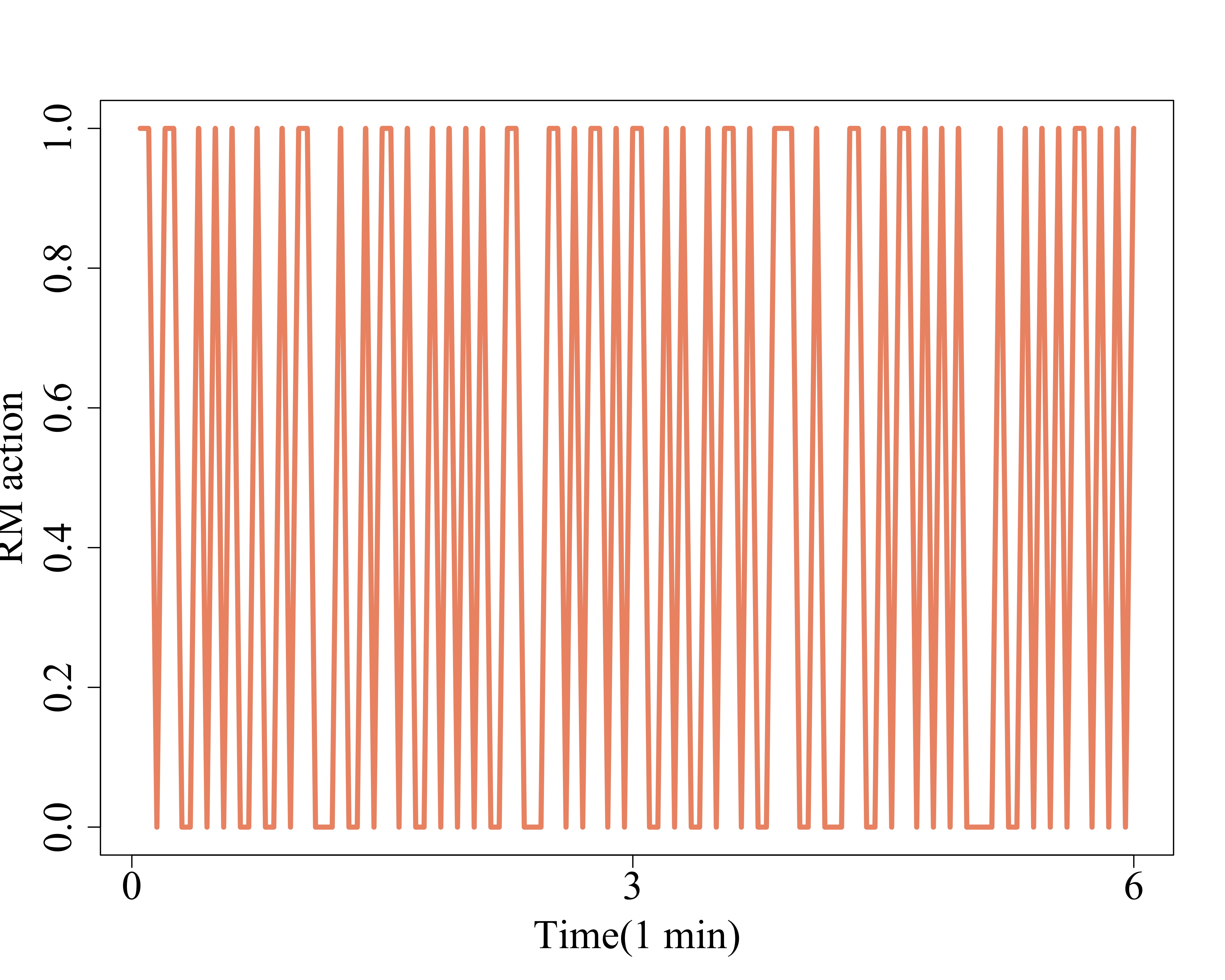}}
  \hspace{0.0in}
  \subfigure[The variable speed limits produced by ES-CTC]{
    \label{DNs2} %% label for first subfigure
    \includegraphics[width=0.2\textwidth]{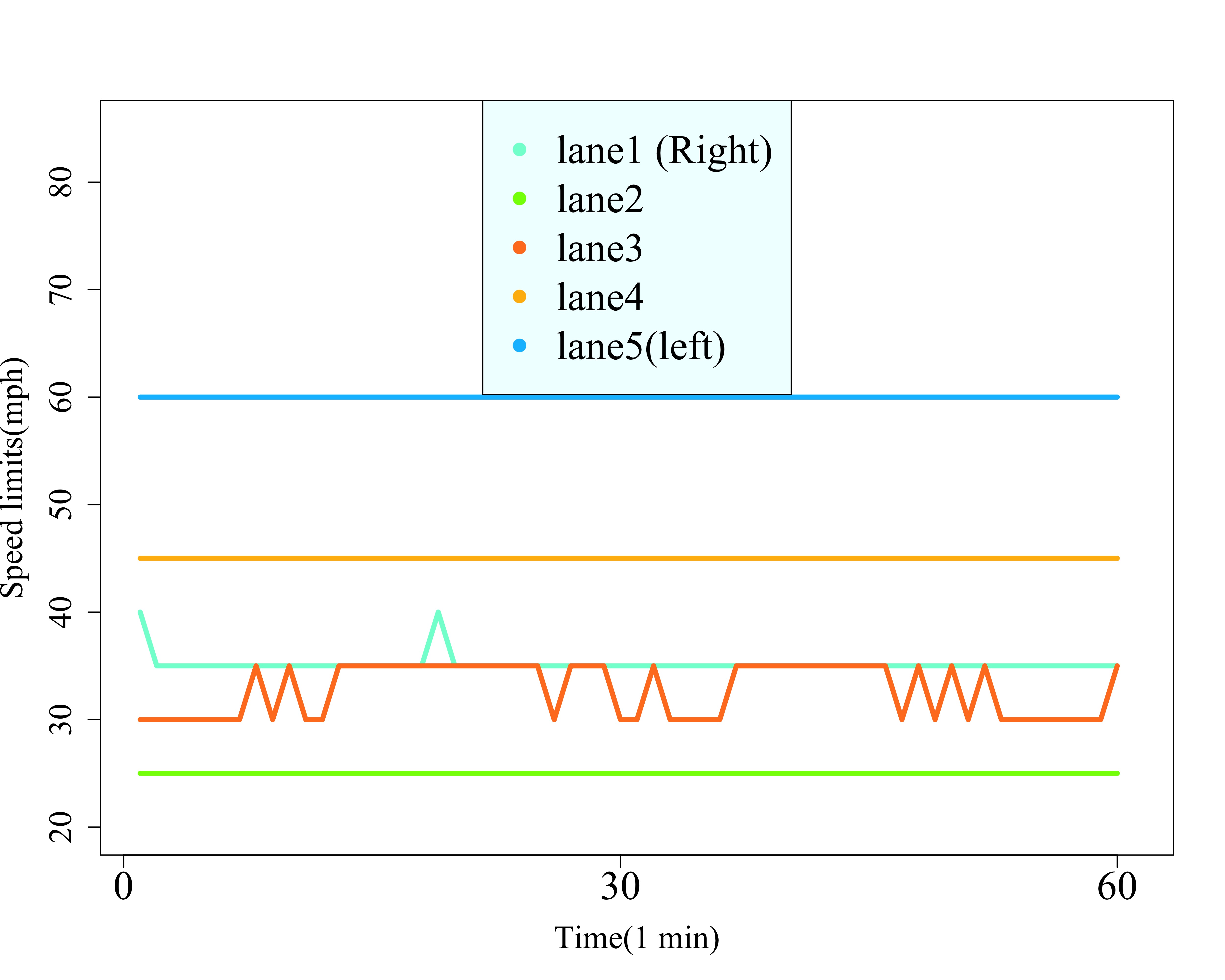}}
  \hspace{0.0in}
    \subfigure[The LCC action produced by ES-CTC]{
    \label{DNs2} %% label for first subfigure
    \includegraphics[width=0.2\textwidth]{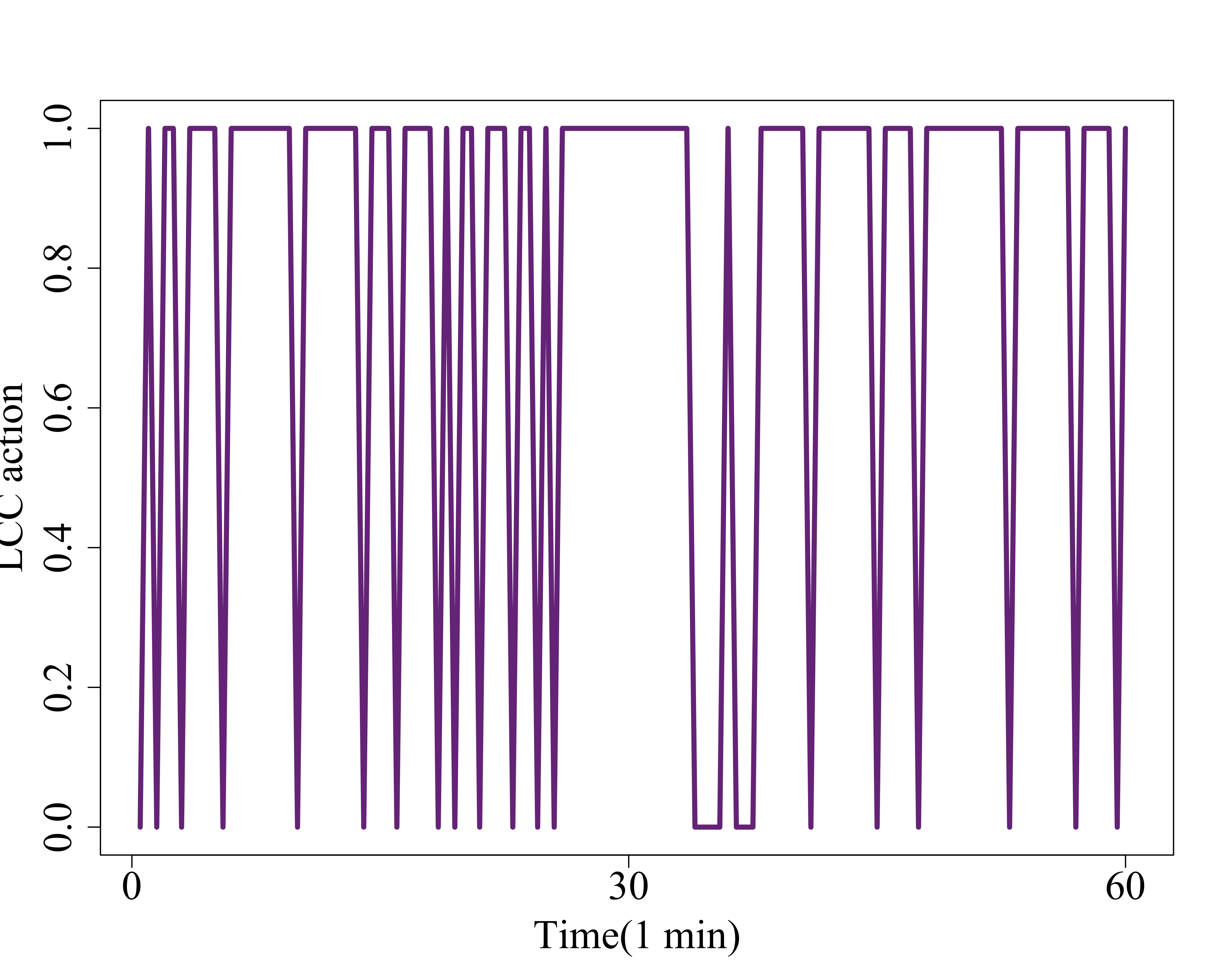}}
  \hspace{0.0in}
  \caption{Visualization of RM, DVSL and LCC actions produced by ES-CTC.}
  \label{policy} %% label for entire figure
\end{figure}

The RM, DVSL and LCC actions of ES-CTC obtained from one simulation are presented in Figure \ref{policy}. The most interesting one is the speed limits produced by DVSL agent. The DVSL agent has learned to always set a maximum speed limit for the leftest lane. it automatically set the left lanes as overtaking lanes. The agents mainly adjusts inflow to the bottleneck by adjusting the speed limits of the right lanes, on-ramp vehicles and vehicles' lane change behaviors. As stated before, the conflicts between vehicles occur mostly in the right lanes. Therefore it is not necessary to decrease the speed limits of left 2 lanes (lane 4 and lane 5). 

\section{Conclusion}
In this paper we have proposed a deep neuroevolutional model for cooperative freeway traffic control. In order to learn the spatial dependence between traffic sensors, the neural networks structure of the model are built upon graph convolutional layer. Our structure allows several traffic control agents with different control cycles work cooperatively to improve the freeway traffic efficiency. Our solution outperforms the state-of-the-art DRL based solutions in terms of improvements in freeway capacity.

Several interesting questions stem from our paper both theoretically and practically, that we plan to study in the future. We aim to extend the approach to large freeway networks and a broader set of dynamic events such as adverse weather and traffic incidents in the future. Another interesting direction we plan to study is the incorporation of more advanced traffic control strategies. In this paper, the most basic graph convolutional network architecture and evolutionary strategy are used. We believe that a more systemic research of architectures and optimization strategies may provide improvements in control performance. 

%would not be acceptable because the text is too small.

%\section{Algorithms and Listings}

%Algorithms and listings are a special kind of figures. Like all illustrations, they should appear floated to the top (preferably) or bottom of the page. However, their caption should appear in the header, left-justified and enclosed between horizontal lines, as shown in Algorithm~\ref{alg:algorithm}. The algorithm body should be terminated with another horizontal line. It is up to the authors to decide whether to show line numbers or not, how to format comments, etc.

%In \LaTeX{} algorithms may be typeset using the {\tt algorithm} and {\tt algorithmic} packages, but you can also use one of the many other packages for the task.  

%\begin{algorithm}[tb]
%\caption{Example algorithm}
%\label{alg:algorithm}
%\textbf{Input}: Your algorithm's input\\
%\textbf{Parameter}: Optional list of parameters\\
%\textbf{Output}: Your algorithm's output
%\begin{algorithmic}[1] %[1] enables line numbers
%\STATE Let $t=0$.
%\WHILE{condition}
%\STATE Do some action.
%\IF {conditional}
%\STATE Perform task A.
%\ELSE
%\STATE Perform task B.
%\ENDIF
%\ENDWHILE
%\STATE \textbf{return} solution
%\end{algorithmic}
%\end{algorithm}

\section*{Acknowledgments}
The work was supported by national natural science foundation of China (61620106002). Any opinions expressed in this paper are solely those of the authors and do
not represent those of the sponsors. The authors would like to thank experienced anonymous
reviewers for their constructive and valuable suggestions for improving the overall quality of
this paper.
%The preparation of these instructions and the \LaTeX{} and Bib\TeX{}
%files that implement them was supported by Schlumberger Palo Alto
%Research, AT\&T Bell Laboratories, and Morgan Kaufmann Publishers.
%Preparation of the Microsoft Word file was supported by IJCAI.  An
%early version of this document was created by Shirley Jowell and Peter
%F. Patel-Schneider.  It was subsequently modified by Jennifer
%Ballentine and Thomas Dean, Bernhard Nebel, Daniel Pagenstecher,
%Kurt Steinkraus, Toby Walsh and Carles Sierra. The current version 
%has been prepared by Marc Pujol-Gonzalez and Francisco Cruz-Mencia.

\appendix

%\section{\LaTeX{} and Word Style Files}\label{stylefiles}

%The \LaTeX{} and Word style files are available on the IJCAI--19
%website, \url{http://www.ijcai19.org}.
%These style files implement the formatting instructions in this
%document.

%The \LaTeX{} files are {\tt ijcai19.sty} and {\tt ijcai19.tex}, and
%the Bib\TeX{} files are {\tt named.bst} and {\tt ijcai19.bib}. The
%\LaTeX{} style file is for version 2e of \LaTeX{}, and the Bib\TeX{}
%style file is for version 0.99c of Bib\TeX{} ({\em not} version
%0.98i). The {\tt ijcai19.sty} style differs from the {\tt
%ijcai18.sty} file used for IJCAI--18.

%The Microsoft Word style file consists of a single file, {\tt
%ijcai19.doc}. This template differs from the one used for
%IJCAI--18.

%These Microsoft Word and \LaTeX{} files contain the source of the
%present document and may serve as a formatting sample.  

%Further information on using these styles for the preparation of
%papers for IJCAI--19 can be obtained by contacting {\tt
%pcchair@ijcai19.org}.

%% The file named.bst is a bibliography style file for BibTeX 0.99c
\bibliographystyle{named}
\bibliography{ijcai19}

\end{document}